\documentclass
[pra,a4paper,superscriptaddress,twocolumn,showpacs,hyperref,nofootinbib]{revtex4}%
\pdfoutput=1
\usepackage{amsfonts}
\usepackage{hyperref}
\usepackage{amsmath}
\usepackage{amssymb}
\usepackage{graphicx}%
\providecommand{\U}[1]{\protect\rule{.1in}{.1in}}

\begin{document}
\title{Nonclassical correlation in NMR quadrupolar systems}
\author{D. O. Soares-Pinto}
\email{diogo.osp@ursa.ifsc.usp.br}
\affiliation{Instituto de F\'{\i}sica de S\~{a}o Carlos, Universidade de S\~{a}o Paulo,
P.O. Box 369, 13560-970 S\~{a}o Carlos, S\~{a}o Paulo, Brazil}
\author{L. C. C\'{e}leri}
\email{lucas.celeri@ufabc.edu.br}
\affiliation{Centro de Ci\^{e}ncias Naturais e Humanas, Universidade Federal do ABC, R.
Santa Ad\'{e}lia 166, 09210-170 Santo Andr\'{e}, S\~{a}o Paulo, Brazil}
\author{R. Auccaise}
\affiliation{Instituto de F\'{\i}sica de S\~{a}o Carlos, Universidade de S\~{a}o Paulo,
P.O. Box 369, 13560-970 S\~{a}o Carlos, S\~{a}o Paulo, Brazil}
\author{F. F. Fanchini}
\affiliation{Instituto de F\'{\i}sica Gleb Wataghin, Universidade Estadual de Campinas,
P.O. Box 6165, 13083-970 Campinas, S\~{a}o Paulo, Brazil}
\author{E. R. deAzevedo}
\affiliation{Instituto de F\'{\i}sica de S\~{a}o Carlos, Universidade de S\~{a}o Paulo,
P.O. Box 369, 13560-970 S\~{a}o Carlos, S\~{a}o Paulo, Brazil}
\author{J. Maziero}
\affiliation{Centro de Ci\^{e}ncias Naturais e Humanas, Universidade Federal do ABC, R.
Santa Ad\'{e}lia 166, 09210-170 Santo Andr\'{e}, S\~{a}o Paulo, Brazil}
\author{T. J. Bonagamba}
\affiliation{Instituto de F\'{\i}sica de S\~{a}o Carlos, Universidade de S\~{a}o Paulo,
P.O. Box 369, 13560-970 S\~{a}o Carlos, S\~{a}o Paulo, Brazil}
\author{R. M. Serra}
\affiliation{Centro de Ci\^{e}ncias Naturais e Humanas, Universidade Federal do ABC, R.
Santa Ad\'{e}lia 166, 09210-170 Santo Andr\'{e}, S\~{a}o Paulo, Brazil}

\begin{abstract}
The existence of quantum correlation (as revealed by quantum discord), other
than entanglement and its role in quantum-information processing (QIP), is a
current subject for discussion. In particular, it has been suggested that this
nonclassical correlation may provide computational speedup for some quantum
algorithms. In this regard, bulk nuclear magnetic resonance (NMR) has been
successfully used as a test bench for many QIP implementations, although it
has also been continuously criticized for not presenting entanglement in most
of the systems used so far. In this paper, we report a theoretical and
experimental study on the dynamics of quantum and classical correlations in an
NMR quadrupolar system. We present a method for computing the correlations
from experimental NMR deviation-density matrices and show that, given the
action of the nuclear-spin environment, the relaxation produces a monotonic
time decay in the correlations. Although the experimental realizations were
performed in a specific quadrupolar system, the main results presented here
can be applied to whichever system uses a deviation-density matrix formalism.

\end{abstract}

\maketitle

\section{Introduction}

Since the birth of quantum information theory (QIT), entanglement has been
considered as a key resource for the processing of information at a quantum
level. However, it is known that quantum correlation may be present even in
separable states. This nonclassical correlation can be quantified, for
example, by the so-called quantum discord \cite{2002_PRL_88_017901}, which is
a measure of a \textquotedblleft gap\textquotedblright\ between quantum- and
classical- information theory. Some other measures for both classical and
quantum correlations\ contained in a bipartite quantum state were proposed in
literature \cite{2001_JPhysA_34_6899, 2003_PRL_90_050401, 2002_PRL_89_180402,
2005_PRA_72_032317, 2008_PRA_77_022301, 2009_ArXiv_0911.5417, Terhal,
DiVincenzo}. Nonclassical correlation may exist in almost all quantum states
\cite{2009_ArXiv_0908.3157}, and it has been theoretically
\cite{2008_PRL_100_050502} and experimentally \cite{2008_PRL_101_200501}
demonstrated that it may provide computational speedup in a model of quantum
computation. Also, it was suggested \cite{2009_ArXiv_0906.3656} that the
speedup of some quantum algorithms may be due to quantum correlation of
separable states.

The study of the nonclassical aspects of a correlated quantum system,
especially the aspect revealed by quantum discord, received a great deal of
attention in recent scientific literature \cite{2003_PRA_67_012320,
2005_PRA_71_062307, 2008_JPhysA_41_205301, 2008_PRL_100_050502,
2008_PRL_100_090502, 2009_PRA_80_044102, 2009_PRL_102_100402,
2009_PRA_79_042325, 2009_PRL_102_250503, 2009_PRA_80_024103,
2009_ArXiv_0911.1096, 2009_ArXiv_0910.5711, Symmetric, Maziero, Unruh}. In
particular, the action of decoherence on this correlation was studied by
taking different kinds of environmental interactions \cite{2009_PRA_80_044102,
2009_PRA_80_024103, 2009_ArXiv_0911.1096,
2009_ArXiv_0910.5711,FanchiniCastelano} into account. Until now, only two
experimental measurements of such a nonclassical correlation have been
performed \cite{2008_PRL_101_200501, 2009_ArXiv_0911.2848}. In Ref.
\cite{2008_PRL_101_200501}, by means of an optical architecture, the authors
implemented the so-called deterministic quantum computation with one qubit
\cite{1998_PRL_81_5672} for the trace estimation of an unitary matrix. In such
a non-universal quantum-information-processing (QIP) model, entanglement is
not a necessary resource for obtaining a computational speedup (in comparison
with the best classical algorithm). The authors of Ref.
\cite{2009_ArXiv_0911.2848} demonstrated, also using an optical setup, the
sudden change in the decay rates of classical and quantum correlations,
theoretically predicted in Ref. \cite{2009_PRA_80_044102}.

Nuclear magnetic resonance (NMR) systems have been extensively used as a
method for implementing and benching test QIP ideas \cite{livro_NMRivan,
livro_suter}. The main feature of the technique is the excellent control of
unitary transformations provided by the use of radio-frequency (rf) pulses,
which result in unique methods for quantum state manipulation
\cite{2005_JChemPhys_122_041101, 2006_PRL_96_170501} and generation of
protocols for processing quantum information \cite{2000_ForPhysik_48_875,
livro_NMRivan, 1998_Nature_396_52}. For example, algorithms such as
Deutsch-Jozsa, Shor, and Grover were successfully tested using NMR systems
\cite{livro_NMRivan, livro_suter}. However, most of these achievements were
performed in bulk samples (bulk NMR) by using the so-called pseudopure states,
for which the existence of entanglement was promptly ruled out
\cite{1999_PRL_83_1054, 2001_PRL_87_047901, 2002_QIC_2_166}. This last fact
has led to the questioning of the quantum nature of NMR implementations for
QIP. On the other hand, as suggested in Refs. \cite{2009_ArXiv_0906.3656,
2002_QIC_2_166}, the existence of quantum correlation, other than entanglement
in NMR systems, may be the reason for the success of most bulk NMR
implementations. Moreover, as mentioned before, quantum correlation of
separable states may provide computational speedup in some tasks.

To our knowledge, here, we present the first demonstration that both quantum
and classical correlations in NMR systems can be determined from
experimentally detectable deviation-density matrices. We use an NMR
quadrupolar system to theoretically and experimentally study the existence of
quantum and classical correlations in a two-logical-qubit composite system.
Our results show that such a nonclassical correlation can easily be created
and can be manipulated at room temperature. We also investigate how these
correlations are degraded by the effect of the environment on the nuclear
spin. In our system, the quantum aspect of correlation decays monotonically
within the relaxation time of the\ deviation matrix. Our experimental
implementations were carried out by using an effective two-qubit
representation of an $^{23}$Na nuclear spin ($I=3/2$) in a lyotropic
liquid-crystal [sodium dodecyl sulfate (SDS)] sample at 26$^{o}$ C. The
achievements presented here, together with the discussion about the role of
quantum correlation (of separable states) in QIP, reinforce the relevance of
the NMR tools in this scenario.

This paper is organized as follows. In Sec. II, we briefly review the
description of the NMR quadrupolar system. Section III is devoted to
presenting the method for determining the quantum and classical correlations
from the experimentally accessible NMR deviation-density matrix. Also in this
section, we present a theoretical model to describe the relaxation dynamics in
our system. The experimental results are presented in Sec. IV; and, in Sec. V,
we present a summary and a discussion.

\section{Description of the System}

In NMR systems, at room temperature, the energy gaps between the levels of the
system $\delta E=\hbar\omega_{L}$ are much smaller than the thermal energy
$\epsilon=\hbar\omega_{L}/2k_{B}T\sim10^{-5}$ (we added, for convenience, a
$1/2$ factor in the definition of $\epsilon$). This implies that a typical
NMR-system density matrix can be written, in the high-temperature expansion
\cite{livro_abragam, livro_NMRivan}, as%
\begin{equation}
\rho\approx\frac{1}{Z}\mathbf{1}+\epsilon\Delta\rho, \label{DM}%
\end{equation}
where $\mathbf{1}$ is the $2^{n}\times2^{n}$ identity matrix, with $n$ being
the number of effective logical qubits in the QIP terminology, $Z\approx2^{n}$
is the system-partition function \cite{livro_NMRivan}, and $\Delta\rho$ is the
traceless deviation-density matrix. In general, the state described in Eq.
(\ref{DM}) is a mixed state that does not possess entanglement
\cite{1999_PRL_83_1054,2001_PRL_87_047901}. However, as we will show in what
follows, it might have quantum correlation (of separable states) that can be
used in QIP protocols \cite{2008_PRL_100_050502, 2008_PRL_101_200501}.

Any manipulation ---such as state preparation, state tomography, qubit
rotations, and so on--- is performed only onto the deviation-matrix
$\Delta\rho$. For example, a sequence of rf pulses, which can be represented
by a unitary transformation $U$, changes the density operator in the following
way:%
\begin{equation}
U\rho U^{\dag}=\frac{1}{2^{n}}\mathbf{1}+\epsilon U\Delta\rho U^{\dag}.
\label{eq.06}%
\end{equation}
By a suitable adjustment of each rf pulse length, phase, and amplitude, very
fine control over the density matrix populations and coherences (diagonal and
off-diagonal elements, respectively) can be achieved. Together with proper
temporal or spatial averaging procedures \cite{2004_RMP_76_1037} and evolution
under spin interactions, the rf pulse can be specially designed to prepare the
system in all two-qubit computational base states as well as its
superpositions, starting from the thermal-equilibrium state (Boltzmann
distribution) \cite{2004_PRA_69_042322,2005_JMagRes_175_226}. The effect of
the environment on the spin system is to induce relaxation in such a way that,
after characteristic times, the populations return to the equilibrium
distribution and the coherences vanish.

The purpose of this paper is to theoretically and experimentally investigate
the presence of quantum and classical correlations and their degradation due
to decoherence for different initial states of a two-qubit representation of
the nuclear spin system. To achieve this goal, we use a spin $I=3/2$
quadrupolar NMR system. In the presence of a static magnetic field, due to the
Zeeman splitting, nuclei with $I>1/2$ can be described by a $(2I+1)$-level
system with equally spaced energy levels. However, such nuclei also possess
quadrupole moments, which interact with the electric-field gradient (EFG)
produced by the charge distribution in their surroundings, the so-called
quadrupolar interaction \cite{livro_NMRivan}. When the Zeeman interaction is
much stronger than the quadrupolar one, the latter can be treated in the
framework of first-order perturbation theory, and the system's Hamiltonian
turns out to be \cite{livro_ernst,livro_apg,livro_slichter}
\begin{equation}
\mathcal{H}=-\hbar\omega_{L}I_{z}+\hbar\omega_{Q}\left(  3{I}_{z}%
^{2}-{\mathbf{I}}^{2}\right)  ,\label{Ham}%
\end{equation}
where $\omega_{L}$ and $\omega_{Q}$ are the Larmor and the quadrupole
frequencies, respectively ($\left\vert \omega_{L}\right\vert \gg\left\vert
\omega_{Q}\right\vert $). The spin-nuclear operator is characterized by its
$z$ component $I_{z}$ and its square modulus ${\mathbf{I}}^{2}$. The first
term of Eq. (\ref{Ham}) describes the Zeeman interaction while the second one
accounts for the static first-order quadrupolar interaction
\cite{livro_ernst,livro_apg,livro_slichter}.%
\begin{figure}
[pt]
\begin{center}
\includegraphics[
natheight=3.844100in,
natwidth=8.552100in,
height=1.5869in,
width=3.3477in
]%
{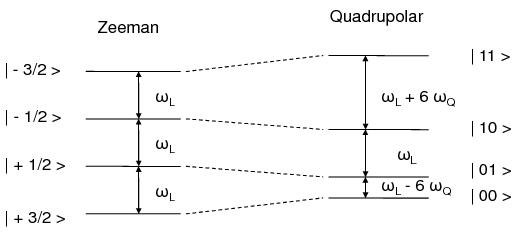}%
\caption{Sketch of the four energy levels in the spin $I=3/2$ system
(characterized by the $z$ component of the nuclear-spin magnetization) for the
Zeeman and the quadrupolar couplings. The indexation of the effective
two-qubit computational basis is also displayed.}%
\end{center}
\end{figure}

This system can be regarded as an effective two-qubit quantum system that may
be used for QIP \cite{2000_ForPhysik_48_875, 2001_ProgNMRSpec_38_325,
2004_RMP_76_1037, livro_NMRivan, livro_suter}. Figure 1 shows a sketch of the
four energy levels in the case $I=3/2$, characterized by the $z$ component of
the nuclear-spin magnetization whose projections are $m=-3/2$, $-1/2$, $1/2$,
$3/2$. Figure 1 also displays the logical indexing of the angular momentum
basis [i.e., the computational basis (for the effective two-qubit system)] as
well as the three single quantum transitions that can be directly detected
after applying a non selective $\left.  \pi\right/  2$ rf pulse onto the
system (equilibrium spectrum). Although the implementation of the qubit in a
quadrupolar system, as considered here, is not so obvious, many experiments
proved that logic gates as well as quantum algorithms can be realized in such
an effective system \cite{2000_JChemPhys_112_6963, 2001_JChemPhys_114_4415}.

\section{Quantum and classical correlations and their dynamics under
decoherence}

Let us now briefly review two proposed measures to quantify the quantum and
classical aspects of the correlation contained in a bipartite quantum system.
The nonclassical correlation in a composed state can be quantified by quantum
discord \cite{2002_PRL_88_017901}, and its classical counterpart is given by
the Henderson-Vedral measure \cite{2001_JPhysA_34_6899, 2003_PRL_90_050401}.
In what follows, we will also discuss a measure of quantum correlation that
can be regarded as a symmetrical version of quantum discord
\cite{2009_ArXiv_0910.5711, Symmetric} and its classical counterpart
\cite{Terhal, DiVincenzo}. A suitable expansion of the correlations measures,
in terms of the experimental measurable deviation matrix [cf. Eq. (\ref{DM})],
will then be obtained. Finally, we present the dynamic equations governing the
evolution of the density operator of a spin $I=3/2$ system under the action of
a typical spin environment in an NMR quadrupolar scenario, using the Redfield
formalism \cite{1957_IBMJResDev_1_19, livro_petruccione}.

\subsection{Measures of correlations}

The mutual information is a measure of the correlation between two random
variables $A$ and $B$ in classical information theory \cite{livro_InfoTheory}%
\begin{equation}
I_{c}(A\text{:}B)=\mathcal{H}(A)+\mathcal{H}(B)-\mathcal{H}(A,B),\label{CMIa}%
\end{equation}
where $\mathcal{H}(X)=-\sum\nolimits_{x}p_{x}\log_{2}p_{x}$ is the Shannon
entropy of the variable $X$ with $p_{x}$ being the probability of variable $X$
assuming the value $x$. By means of the Bayes rule, we can rewrite Eq.
(\ref{CMIa}) in an equivalent form%
\begin{equation}
J_{c}(A\text{:}B)=\mathcal{H}(A)-\mathcal{H}(A|B),\label{CMIb}%
\end{equation}
with $\mathcal{H}(A|B)=-\sum_{a,b}p_{a,b}\log_{2}p_{a|b}$ ($p_{a|b}=\left.
p_{a,b}\right/  p_{b}$) being the conditional entropy, which represents the
lack of knowledge of variable $A$ when we know (by means of a measurement) the
value of variable $B$. It is clear that, in the classical case, we always have
$I_{c}(A$:$B)-J_{c}(A$:$B)=0$.

In the realm of QIT, while the extension of Eq. (\ref{CMIa}) for a bipartite
quantum state ($\rho_{AB}$) is trivially obtained as \cite{livro_nielsen,
livro_benenti, livro_vedral}%
\begin{equation}
\mathcal{I}(\rho_{A:B})=S(\rho_{A})+S(\rho_{B})-S(\rho_{AB})\text{,}%
\label{QMI}%
\end{equation}
the extension of Eq. (\ref{CMIb}) is not so straightforward. Here
$S(\rho)=-\operatorname*{Tr}(\rho\log_{2}\rho)$ is the von Neumann entropy,
and $\rho_{A(B)}=\operatorname*{Tr}_{B(A)}(\rho_{AB})$ is the reduced-density
operator of partition $A$($B$). The quantum mutual information given in Eq.
(\ref{QMI}) is a measure of the total correlation (including the classical and
the quantum ones) contained in a bipartite quantum system
\cite{2005_PRA_72_032317, 2006_PRA_74_042305}. The quantum extensions of Eqs.
(\ref{CMIa}) and (\ref{CMIb}) are not equivalent, in general; and this
inequivalence relies on the distinct nature of both quantum and classical
measurements. While the classical measurement can be chosen such that it does
not disturb the system to be measured; in the quantum case, the measurement
process may affect the system in a fundamental manner. This observation leads
Ollivier and Zurek \cite{2002_PRL_88_017901} to propose a measure of quantum
correlation, named quantum discord, given by the difference%
\begin{equation}
\mathcal{D}(\rho_{AB})\equiv\mathcal{I}(\rho_{A:B})-\max_{\left\{  \Pi_{j}%
^{B}\right\}  }\mathcal{J}(\rho_{A:B}),\label{Disc}%
\end{equation}
where $\mathcal{J}(\rho_{A:B})=S(\rho_{A})-S_{\left\{  \Pi_{j}^{B}\right\}
}\left(  \rho_{A|B}\right)  $, with $S_{\left\{  \Pi_{j}^{B}\right\}  }\left(
\rho_{A|B}\right)  =\sum_{j}q_{j}S\left(  \rho_{A}^{j}\right)  $ being a
quantum extension of the classical conditional entropy $\mathcal{H}(A|B)$. The
reduced state of partition $A$ ($\rho_{A}^{j}$), after the measurement
$\Pi_{j}^{B}$ is performed on partition $B$, is given by $\rho_{A}%
^{j}=\operatorname*{Tr}\nolimits_{B}\left.  \left\{  \left(  \mathbf{1}%
_{A}\otimes\Pi_{j}^{B}\right)  \rho_{AB}\left(  \mathbf{1}_{A}\otimes\Pi
_{j}^{B}\right)  \right\}  \right/  q_{j}$, with $q_{j}=\operatorname*{Tr}%
_{AB}\left[  \left(  \mathbf{1}_{A}\otimes\Pi_{j}^{B}\right)  \rho
_{AB}\right]  $ ($\mathbf{1}_{A}$ is the identity operator for partition $A$).
The quantum discord Eq. (\ref{Disc}) is computed by an extremization procedure
over all possible complete sets of projective measurements $\left\{  \Pi
_{j}^{B}\right\}  $ over subsystem $B$.

Due to the distinct nature of quantum and classical correlations, we can
assume that both correlations add up in a simple way to give the quantum
mutual information. Therefore, the classical counterpart of the quantum
discord may be defined simply as \cite{2009_PRA_80_044102,
2009_ArXiv_0910.5711}%
\begin{equation}
\mathcal{C}(\rho_{AB})\equiv\mathcal{I}(\rho_{A:B})-\mathcal{D}(\rho_{AB}),
\label{Class}%
\end{equation}
that is identical to the Henderson-Vedral definition of classical correlation
\cite{2001_JPhysA_34_6899, 2003_PRL_90_050401}.

In some circumstances the quantum discord, Eq. (\ref{Disc}), and also its
classical counterpart, Eq. (\ref{Class}), may be asymmetric with respect to
the choice of the partition to be measured (see Refs.
\cite{2009_ArXiv_0910.5711, Symmetric} for a related discussion concerning
symmetric measures of correlations). Since the states considered in this paper
are, in general, asymmetric by an interchange between the two partitions we
will use symmetrized versions for the measures of both quantum and classical
correlations. Alternatively, the classical correlation in a composite
bipartite system can be expressed as the \ maximum classical mutual
information\ that is obtained by local measurements over both partitions of a
composite state \cite{Terhal, DiVincenzo}%
\begin{equation}
\mathcal{K}(\rho_{AB})\equiv\underset{\left\{  \Pi_{i}^{A}\otimes\Pi_{j}%
^{B}\right\}  }{\max}\left[  I_{c}(A\text{:}B)\right]  ,\label{SCC}%
\end{equation}
where $I_{c}(A$:$B)$ is the classical mutual information defined in Eq.
(\ref{CMIa}) for the probability distributions that result from the
quantum-measurement process. The extremization in Eq. (\ref{SCC}) is taken
over the set of local projective measurements $\left\{  \Pi_{i}^{A}\otimes
\Pi_{j}^{B}\right\}  $ over both subsystems. Since the quantum mutual
information quantifies the total correlation, a symmetrized measure of the
quantum correlation can be defined as \cite{2009_ArXiv_0910.5711, Symmetric}%
\begin{equation}
\mathcal{Q}(\rho_{AB})\equiv\mathcal{I}(\rho_{A:B})-\mathcal{K}(\rho
_{AB})\text{.}\label{SQC}%
\end{equation}
For composite states of two qubits with maximally mixed marginals the quantum
discord Eq. (\ref{Disc}) is identical to its symmetrized version Eq.
(\ref{SQC}), i.e., $\mathcal{D}(\rho_{AB})=\mathcal{Q}(\rho_{AB})$ [and also
$\mathcal{K}(\rho_{AB})=\mathcal{C}(\rho_{AB})$]. However, it is not true in
general \cite{2009_ArXiv_0910.5711, Symmetric}. We can regard Eq. (\ref{SQC})
as a symmetrical version of the quantum discord. Indeed such a symmetrical
quantifier also reveals a departure between the quantum and the classical
versions of information theory.

\subsection{Correlations and the deviation matrix}

For our purposes, the initial NMR density matrix can be written in terms of
the deviation matrix as%
\begin{equation}
\rho=\frac{\mathbf{1}}{4}+\epsilon\Delta\rho.\label{IS}%
\end{equation}
The deviation matrix, $\Delta\rho$, can be experimentally reconstructed using
a set of rf pulses and readouts [quantum-state tomography (QST)]
\cite{livro_NMRivan, livro_suter}. The parameter $\epsilon$ may be estimated
from the Zeeman and thermal-energy ratios (in our experiment, at room
temperature, $\epsilon\ll1$). Since all correlations present in the state
$\rho$ comes from the deviation matrix $\Delta\rho$, it is convenient to
express the measures of correlations as functions of $\Delta\rho$. To do this,
we will expand the von Neumann's entropy in powers of the small parameter
$\epsilon$ as follows:%

\begin{equation}
S\left(  \rho\right)  =2\left(  1-\frac{\epsilon^{2}}{\ln2}\operatorname*{Tr}%
\Delta\rho^{2}\right)  +\cdots\text{,}\label{Entro}%
\end{equation}
where we used the fact that $\operatorname*{Tr}\Delta\rho=0$. The
reduced-density operators reads $\rho_{A(B)}=\operatorname*{Tr}%
\nolimits_{B(A)}\rho=\left.  \mathbf{1}_{A(B)}\right/  2+\epsilon\Delta
\rho_{A(B)}$, with $\Delta\rho_{A(B)}=\operatorname*{Tr}_{B(A)}\Delta\rho$. We
observe that because $\operatorname*{Tr}\Delta\rho=0$ we have
$\operatorname*{Tr}\Delta\rho_{A}=\operatorname*{Tr}\Delta\rho_{B}=0$. Thus
the marginal entropies are straightforwardly obtained as%
\begin{equation}
S\left(  \rho_{A(B)}\right)  =1-\frac{\epsilon^{2}}{\ln2}\operatorname*{Tr}%
\Delta\rho_{A(B)}^{2}+\cdots\text{.}\label{EntroRed}%
\end{equation}
By replacing Eqs. (\ref{Entro}) and (\ref{EntroRed}) into Eq. (\ref{QMI}) and
by disregarding terms containing high powers in $\epsilon$, we obtain%
\begin{equation}
\mathcal{I}\left(  \rho\right)  \approx\frac{\epsilon^{2}}{\ln2}\left(
2\operatorname*{Tr}\Delta\rho^{2}-\operatorname*{Tr}\Delta\rho_{A}%
^{2}-\operatorname*{Tr}\Delta\rho_{B}^{2}\right)  .\label{QMI_DM}%
\end{equation}

To compute the classical correlation we must obtain the measured density
operator, which is given by
\begin{align}
\eta &  =\sum_{i,j}\left(  \Pi_{i}^{A}\otimes\Pi_{j}^{B}\right)  \rho\left(
\Pi_{i}^{A}\otimes\Pi_{j}^{B}\right) \nonumber\\
&  \equiv\frac{\mathbf{1}}{4}+\epsilon\Delta\eta\text{,} \label{DM_Proj}%
\end{align}
where we defined the measured deviation matrix as $\Delta\eta\equiv\sum
_{i,j}\left(  \Pi_{i}^{A}\otimes\Pi_{j}^{B}\right)  \Delta\rho\left(  \Pi
_{i}^{A}\otimes\Pi_{j}^{B}\right)  $. For a two-qubit system, the complete set
of projective measurements is given by$\ \left\{  \Pi_{j}^{k}=\left\vert
\Theta_{j}^{k}\right\rangle \left\langle \Theta_{j}^{k}\right\vert \text{,
}j=\parallel,\perp\text{, }k=A,B\right\}  $, where $\left\vert \Theta
_{\parallel}^{k}\right\rangle \equiv\cos(\theta_{k})\left\vert 0_{k}%
\right\rangle +e^{i\phi_{k}}\sin(\theta_{k})\left\vert 1_{k}\right\rangle $
and $\left\vert \Theta_{\perp}^{k}\right\rangle \equiv e^{-i\phi_{k}}%
\sin(\theta_{k})\left\vert 0_{k}\right\rangle -\cos(\theta_{k})\left\vert
1_{k}\right\rangle $ with $0\leq\theta_{k}\leq\pi$ and $0\leq\phi_{k}\leq2\pi
$. $\left\{  \left\vert 0_{k}\right\rangle ,\left\vert 1_{k}\right\rangle
\right\}  $ is the computational basis of the logical qubit of partition $k$.
We note that the correlations quantifiers, presented here, have the same
values for simultaneous or successive measurements performed on each partition.

The same reasoning that results in Eq. (\ref{QMI_DM}) leads us to the
following expression, in terms of the measured deviation matrix, for the
mutual information of the measured state,%
\begin{equation}
I_{c}\left(  \eta\right)  \approx\frac{\epsilon^{2}}{\ln2}\left[
2\operatorname*{Tr}\Delta\eta^{2}-\operatorname*{Tr}\left(  \Delta\eta
_{A}\right)  ^{2}-\operatorname*{Tr}\left(  \Delta\eta_{B}\right)
^{2}\right]  ,\label{QMMI}%
\end{equation}
and, thus, for the classical correlation,%
\begin{equation}
\mathcal{K}\left(  \rho\right)  \approx\max_{\left\{  \Pi_{i}^{A}\otimes
\Pi_{j}^{B}\right\}  }I_{c}\left(  \eta\right)  ,\label{CC}%
\end{equation}
where $\Delta\eta_{A(B)}=\operatorname*{Tr}_{B(A)}\Delta\eta$ is the reduced
measured deviation matrix of partition $A$($B$). The quantum correlation in
the composed two-qubit system can be directly computed from Eqs.
(\ref{QMI_DM}) and (\ref{CC}) as $\mathcal{Q}(\rho)=\mathcal{I}\left(
\rho\right)  -\mathcal{K}\left(  \rho\right)  $.

\subsection{Action of the environment on the deviation matrix}%

\begin{figure}
[pt]
\begin{center}
\includegraphics[
natheight=5.864300in,
natwidth=12.927200in,
height=1.8931in,
width=2.7466in
]%
{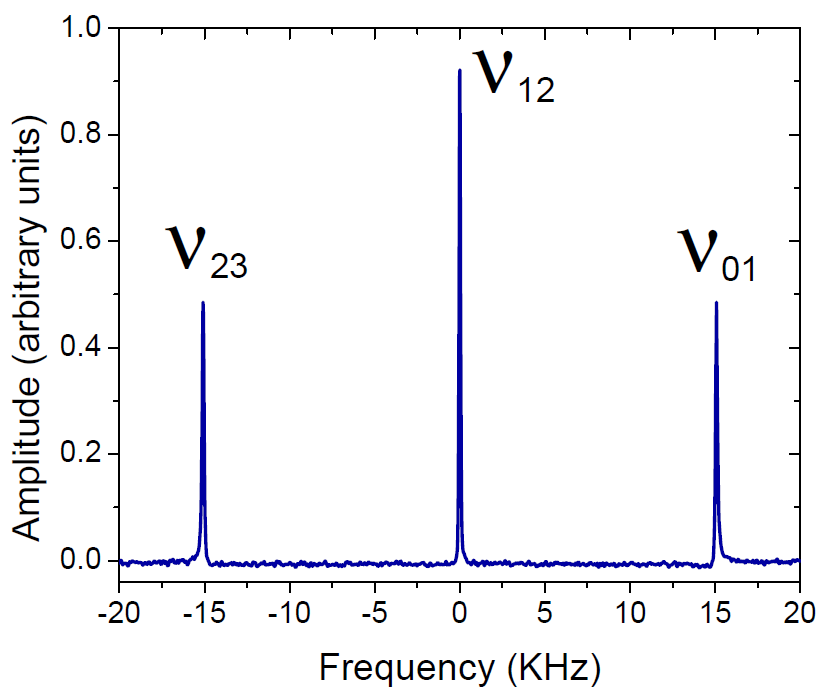}%
\caption{(Color online) $^{23}$Na NMR equilibrium spectrum of the oriented
liquid-crystal SDS sample. The frequencies $\nu_{01}$, $\nu_{12}$, and
$\nu_{23}$ correspond to the transitions $\left\vert 00\right\rangle
\leftrightarrow\left\vert 01\right\rangle $, $\left\vert 01\right\rangle
\leftrightarrow\left\vert 10\right\rangle $, and $\left\vert 10\right\rangle
\leftrightarrow\left\vert 11\right\rangle $, respectively.}%
\end{center}
\end{figure}

As commented on Sec. II, the quadrupolar coupling emerges from the interaction
of the nuclear quadrupole moment with the EFG generated by the surrounding
environment. Internal molecular or atomic motions cause random fluctuations in
the EFG, which introduces noise into the system, which leads to the relaxation
process that causes decoherence and energy dissipation. The dependence of the
random EFG fluctuations on the molecular motions is established by the
spectral densities that encode the motion characteristics, such as geometry
and correlation times \cite{LipariSzabo}.
\begin{figure}
[pt]
\begin{center}
\includegraphics[
natheight=3.427200in,
natwidth=9.041600in,
height=1.2609in,
width=3.282in
]%
{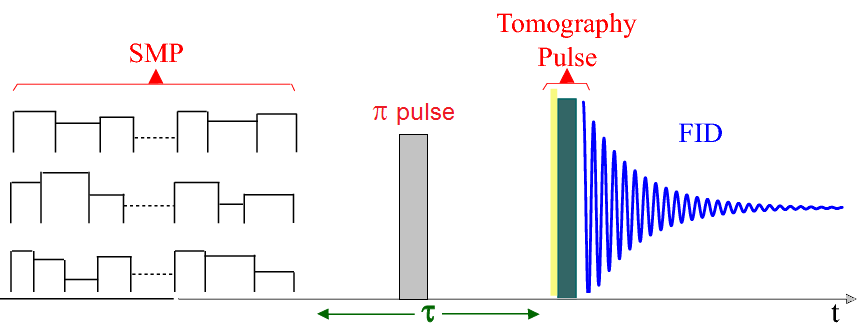}%
\caption{(Color online) Scheme of the pulse sequence used in the experimental
procedure. The first step is the state preparation through a sequence of SMPs
\cite{2002_JChemPhys_116_7599}. Next, we leave the system to evolve only under
the action of the decoherent environment during a variable period of time
$\tau$. The last step is the nonselective tomography pulse and the observation
of the free induction decay (FID) signal, which enable us to reconstruct the
deviation-density matrix $\Delta\rho$ through QST
\cite{2007_JChemPhys_126_154506}. The $\pi$ pulse was added in the middle of
the free evolution period to refocus the $B_{0}$-field inhomogeneities.}%
\end{center}
\end{figure}

Our experimental implementations were carried-out using $^{23}$Na nuclear
spins ($I=3/2$) in a lyotropic liquid crystal at 26$^{o}$ C. In this system,
the relaxation of the $^{23}$Na nuclear spin can be well described by
considering a pure quadrupolar mechanism (the relaxation is produced
exclusively by the EFG fluctuations), so it can be represented by three
reduced spectral densities $J_{0}$, $J_{1}$, and $J_{2}$ at the Larmor
frequency $\omega_{L}$. The specific model that relates the molecular motions
and the reduced spectral densities can be found in Ref.
\cite{2008_JMagRes_192_17}. By applying the Redfield formalism
\cite{livro_petruccione} to our system, enables us to obtain the dynamic
evolution of the deviation-density matrix elements $\Delta\rho_{ij}$ ($i$, $j$
running from $0$ to $3$) as \cite{2008_JMagRes_192_17}%
\begin{subequations}
\begin{align}
\Delta\rho_{01}\left(  t\right)   &  =\frac{1}{2}\left[  \Delta\rho_{01}%
^{0}+\Delta\rho_{23}^{0}\right.  \nonumber\\
&  \left.  +\left(  \Delta\rho_{01}^{0}-\Delta\rho_{23}^{0}\right)
\operatorname*{e}\nolimits^{-2CJ_{2}t}\right]  \operatorname*{e}%
\nolimits^{-C\left(  J_{0}+J_{1}\right)  t},\label{Redeqaa}%
\end{align}%
\begin{align}
\Delta\rho_{02}\left(  t\right)   &  =\frac{1}{2}\left[  \Delta\rho_{02}%
^{0}+\Delta\rho_{13}^{0}\right.  \nonumber\\
&  \left.  +\left(  \Delta\rho_{02}^{0}-\Delta\rho_{13}^{0}\right)
\operatorname*{e}\nolimits^{-2CJ_{1}t}\right]  \operatorname*{e}%
\nolimits^{-C\left(  J_{0}+J_{2}\right)  t},\label{Redeqbb}%
\end{align}%
\begin{align}
\Delta\rho_{13}\left(  t\right)   &  =\frac{1}{2}\left[  \Delta\rho_{02}%
^{0}+\Delta\rho_{13}^{0}\right.  \nonumber\\
&  \left.  -\left(  \Delta\rho_{02}^{0}-\Delta\rho_{13}^{0}\right)
\operatorname*{e}\nolimits^{-2CJ_{1}t}\right]  \operatorname*{e}%
\nolimits^{-C\left(  J_{0}+J_{2}\right)  t},\label{Redeqcc}%
\end{align}%
\begin{align}
\Delta\rho_{23}\left(  t\right)   &  =\frac{1}{2}\left[  \Delta\rho_{01}%
^{0}+\Delta\rho_{23}^{0}\right.  \nonumber\\
&  \left.  -\left(  \Delta\rho_{01}^{0}-\Delta\rho_{23}^{0}\right)
\operatorname*{e}\nolimits^{-2CJ_{2}t}\right]  \operatorname*{e}%
\nolimits^{-C\left(  J_{0}+J_{1}\right)  t},\\
\Delta\rho_{03}\left(  t\right)   &  =\Delta\rho_{03}^{0}\operatorname*{e}%
\nolimits^{-C\left(  J_{1}+J_{2}\right)  t},\\
\Delta\rho_{12}\left(  t\right)   &  =\Delta\rho_{12}^{0}\operatorname*{e}%
\nolimits^{-C\left(  J_{1}+J_{2}\right)  t},\\
\Delta\rho_{00}\left(  t\right)   &  =3-\frac{1}{4}\left[  R_{1}%
\operatorname*{e}\nolimits^{-2C\left(  J_{1}+J_{2}\right)  t}\right.
\nonumber\\
&  \left.  -R_{2}\operatorname*{e}\nolimits^{-2CJ_{2}t}-R_{3}\operatorname*{e}%
\nolimits^{-2CJ_{1}t}\right]  ,\\
\Delta\rho_{11}\left(  t\right)   &  =1+\frac{1}{4}\left[  R_{1}%
\operatorname*{e}\nolimits^{-2C\left(  J_{1}+J_{2}\right)  t}\right.
\nonumber\\
&  \left.  +R_{2}\operatorname*{e}\nolimits^{-2CJ_{2}t}-R_{3}\operatorname*{e}%
\nolimits^{-2CJ_{1}t}\right]  ,\\
\Delta\rho_{22}\left(  t\right)   &  =-1+\frac{1}{4}\left[  R_{1}%
\operatorname*{e}\nolimits^{-2C\left(  J_{1}+J_{2}\right)  t}\right.
\nonumber\\
&  \left.  -R_{2}\operatorname*{e}\nolimits^{-2CJ_{2}t}+R_{3}\operatorname*{e}%
\nolimits^{-2CJ_{1}t}\right]  ,\\
\Delta\rho_{33}\left(  t\right)   &  =-3-\frac{1}{4}\left[  R_{1}%
\operatorname*{e}\nolimits^{-2C\left(  J_{1}+J_{2}\right)  t}\right.
\nonumber\\
&  \left.  +R_{2}\operatorname*{e}\nolimits^{-2CJ_{2}t}+R_{3}\operatorname*{e}%
\nolimits^{-2CJ_{1}t}\right]  .\label{Redeqf}%
\end{align}

In Eqs. (\ref{Redeqaa})-(\ref{Redeqf}), the superscript $0$ refers to the
initial value of each deviation-matrix element and $R_{i}$ ($i=1$, $2$, $3$)
are constant coefficients. The labels of the deviation-matrix elements
$\left\{  0,1,2,3\right\}  $ refer to the computational basis of the effective
two-qubit system ordered in the usual way $\left\{  \left\vert 00\right\rangle
,\left\vert 01\right\rangle ,\left\vert 10\right\rangle ,\left\vert
11\right\rangle \right\}  .$ The parameter $C$ is proportional to the
quadrupolar coupling $\omega_{Q}$ and can be obtained from the equilibrium NMR
spectrum (displayed in Fig. 2) \cite{JCP85_6282,2008_JMagRes_192_17}. To fully
describe the system relaxation, it is also necessary to determine the spectral
densities $J_{i}$ ($i=0$, $1$, $2$) as well as the coefficients $R_{j}$
($j=1$, $2$, $3$). This can be achieved by preparing an initial state by using
the technique of strongly modulated pulses (SMP)
\cite{2002_JChemPhys_116_7599, 2005_JChemPhys_12_2214108,
2007_JChemPhys_126_154506}, together with temporal averaging, where all
$\Delta\rho$ coherences are non zero (the full superposition state), by
letting it evolves under the action of relaxation during a time period $\tau$,
and then by measuring each $\Delta\rho$ element using QST
\cite{2007_JChemPhys_126_154506}. By repeating this procedure for different
values of $\tau$, the relaxation dynamics for each $\Delta\rho$ element is
experimentally measured. In Fig. 3, we depict the pulse-sequence scheme of the
experimental procedure used. We note that Eqs. (\ref{Redeqaa})-(\ref{Redeqf})
are valid for whichever initial state of the system, with the asymptotic state
being the equilibrium one, $\rho=\mathbf{1}/4+\epsilon2I_{Z}$.

Equations (\ref{Redeqaa})-(\ref{Redeqf}) can be combined to provide
single-exponential functions in the following way \cite{2008_JMagRes_192_17}%
\end{subequations}
\begin{subequations}
\begin{align}
\Delta\rho_{01}+\Delta\rho_{23} &  =\left(  \Delta\rho_{01}^{0}+\Delta
\rho_{23}^{0}\right)  e^{-C(J_{0}+J_{1})t},\label{SexpA}\\
\Delta\rho_{02}-\Delta\rho_{13} &  =\left(  \Delta\rho_{02}^{0}+\Delta
\rho_{13}^{0}\right)  e^{-C(J_{0}+J_{2})t},\\
\Delta\rho_{12} &  =\Delta\rho_{12}^{0}e^{-C(J_{1}+J_{2})t},\label{SexpC}%
\end{align}%
\begin{align}
\Delta\rho_{00}+\Delta\rho_{11}-\Delta\rho_{22}-\Delta\rho_{33} &
=8+R_{2}e^{-2CJ_{2}t},\label{SexpD}\\
-\Delta\rho_{00}+\Delta\rho_{11}+\Delta\rho_{22}-\Delta\rho_{33} &
=R_{1}e^{-2C\left(  J_{1}+J_{2}\right)  t},\\
\Delta\rho_{00}-\Delta\rho_{11}+\Delta\rho_{22}-\Delta\rho_{33} &
=4+R_{3}e^{-2CJ_{1}t}.\label{SexpF}%
\end{align}
For convenience, in Eqs. (\ref{SexpA})-(\ref{SexpF}) we have suppressed the
temporal dependence of the deviation-matrix elements $\Delta\rho_{ij}(t).$
Thus, by fitting the experimental evolution of the deviation matrix elements,
it is possible to determine the mean values of the reduced spectral densities
$J_{i}$ by using Eqs. (\ref{SexpA})-(\ref{SexpC}) and to determine the
parameters $R_{j}$ by using Eqs. (\ref{SexpD})-(\ref{SexpF}). We also checked
the consistency of the experimental data with the adopted relaxation model by
comparing the values of $J_{1}$ and $J_{2}$ obtained from Eqs. (\ref{SexpA}%
)-(\ref{SexpC}) and Eqs. (\ref{SexpD})-(\ref{SexpF}) (see Ref.
\cite{2008_JMagRes_192_17} for specific details).

The relaxation dynamics of the system can also be described by the time
dependence of the mean value of $I_{z}$ and $I_{x,y}$ [$\langle I_{z}%
\rangle(t)$ -- longitudinal relaxation, $\langle I_{x,y}\rangle(t)$ --
transverse relaxation], which can be calculated by using Eqs.(\ref{Redeqaa}%
)-(\ref{Redeqf}) as%
\end{subequations}
\begin{align}
\langle I_{z}\rangle(t)-\langle I_{z}\rangle_{T} &  =\frac{1}{2}\left\{
3\Delta\rho_{00}+\Delta\rho_{11}-\Delta\rho_{22}-3\Delta\rho_{33}\right\}
-5\nonumber\\
&  =5+\frac{1}{2}\left\{  2R_{2}e^{-2CJ_{2}t}+R_{3}e^{-2CJ_{1}t}\right\}
,\label{Relax1}%
\end{align}
and%
\begin{align}
\langle I_{x}\rangle(t)+i\langle I_{y}\rangle(t) &  =2\Delta\rho
_{21}\nonumber\\
&  =\Delta\rho_{21}^{0}e^{-C(J_{1}+J_{2})t}\text{.}\label{Relax2}%
\end{align}
In Eq. (\ref{Relax1}), $\langle I_{z}\rangle_{T}$ represents $\langle
I_{z}\rangle$ for the thermal-equilibrium state. The longitudinal relaxation
is then characterized by the time constants $\tau_{L_{1}}=(2CJ_{1})^{-1}$ and
$\tau_{L_{2}}=(2CJ_{2})^{-1}$, and the transverse relaxation is characterized
by the time constant $\tau_{T}=\left[  C(J_{1}+J_{2})\right]  ^{-1}$
\cite{Hubbard,Bull}.

It is important to note that, in our effective two-qubit representation of the
four-level system, the environment acts globally (i.e., it acts simultaneously
on both logical qubits). Such global action of the environment can be observed
in the form of Eqs. (\ref{Redeqaa})-(\ref{Redeqf}), where we do not have
distinct spectral densities for each qubit. We recall that we can consider
this $I=3/2$ system as an effective two-qubit one since we can manipulate the
transitions between the energy levels of these logical qubits (i.e., the
nuclear transitions of the quadrupolar spin) just as it is performed in the
case of physical qubits \cite{2005_JMagRes_175_226}.%

\begin{figure}
[pt]
\begin{center}
\includegraphics[
natheight=12.572600in,
natwidth=9.385800in,
height=3.9271in,
width=2.9386in
]%
{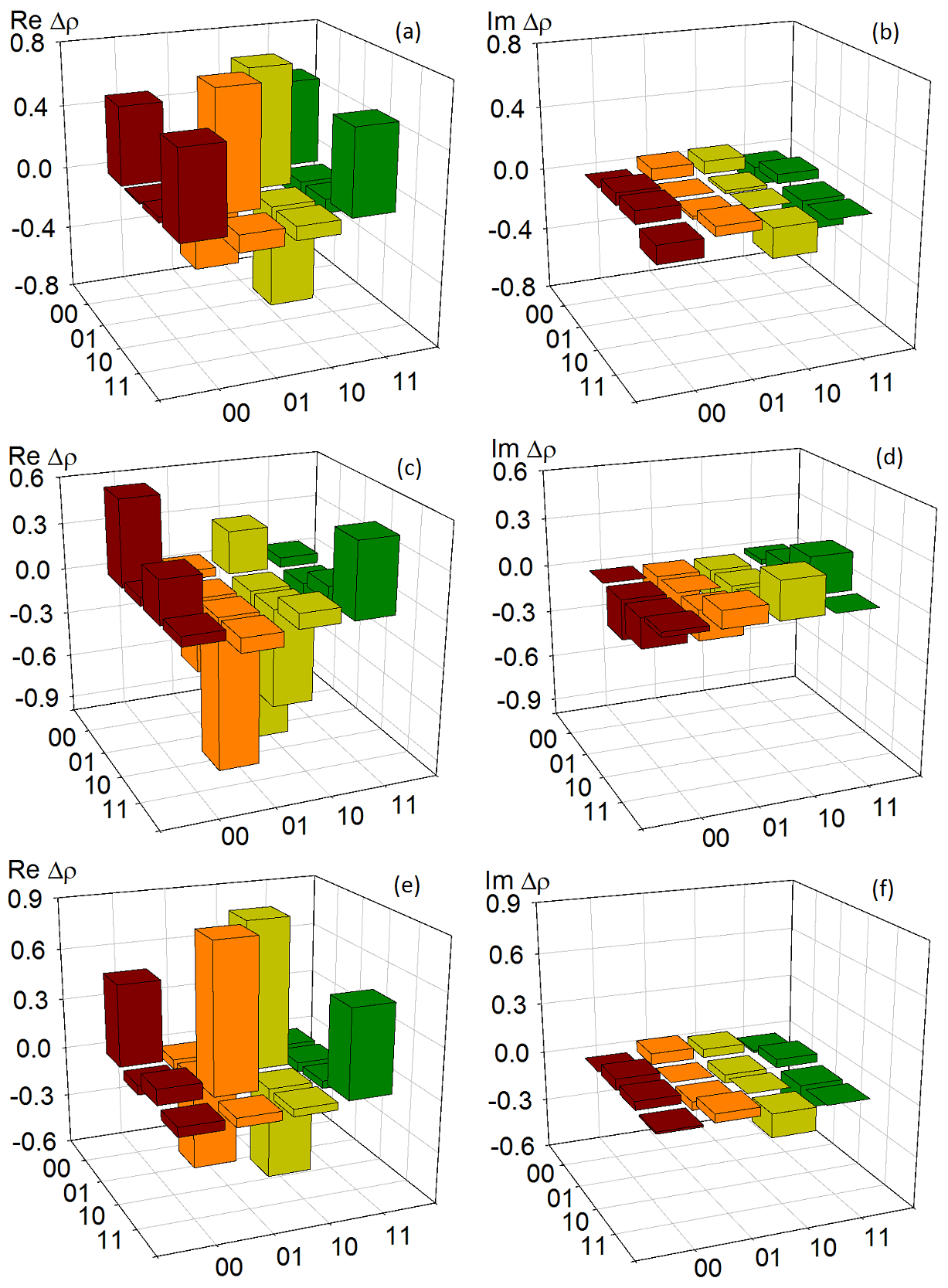}%
\caption{(Color online) Bar representation of the QST experimental data for
the prepared deviation matrix: (a) real and (b) imaginary parts of the
equivalent to the $\left\vert X_{\text{random}}\right\rangle $ pseudopure
state; (c) real and (d) imaginary parts of the equivalent to the $\left\vert
\Phi^{-}\right\rangle $ Bell-basis pseudopure state; (e) real and (f)
imaginary parts of the equivalent to the $\left\vert \Phi^{+}\right\rangle $
Bell-basis pseudopure state. }%
\end{center}
\end{figure}

\section{Experimental results}%

\begin{figure}
[pt]
\begin{center}
\includegraphics[
natheight=12.353900in,
natwidth=9.375400in,
height=3.864in,
width=2.9386in
]%
{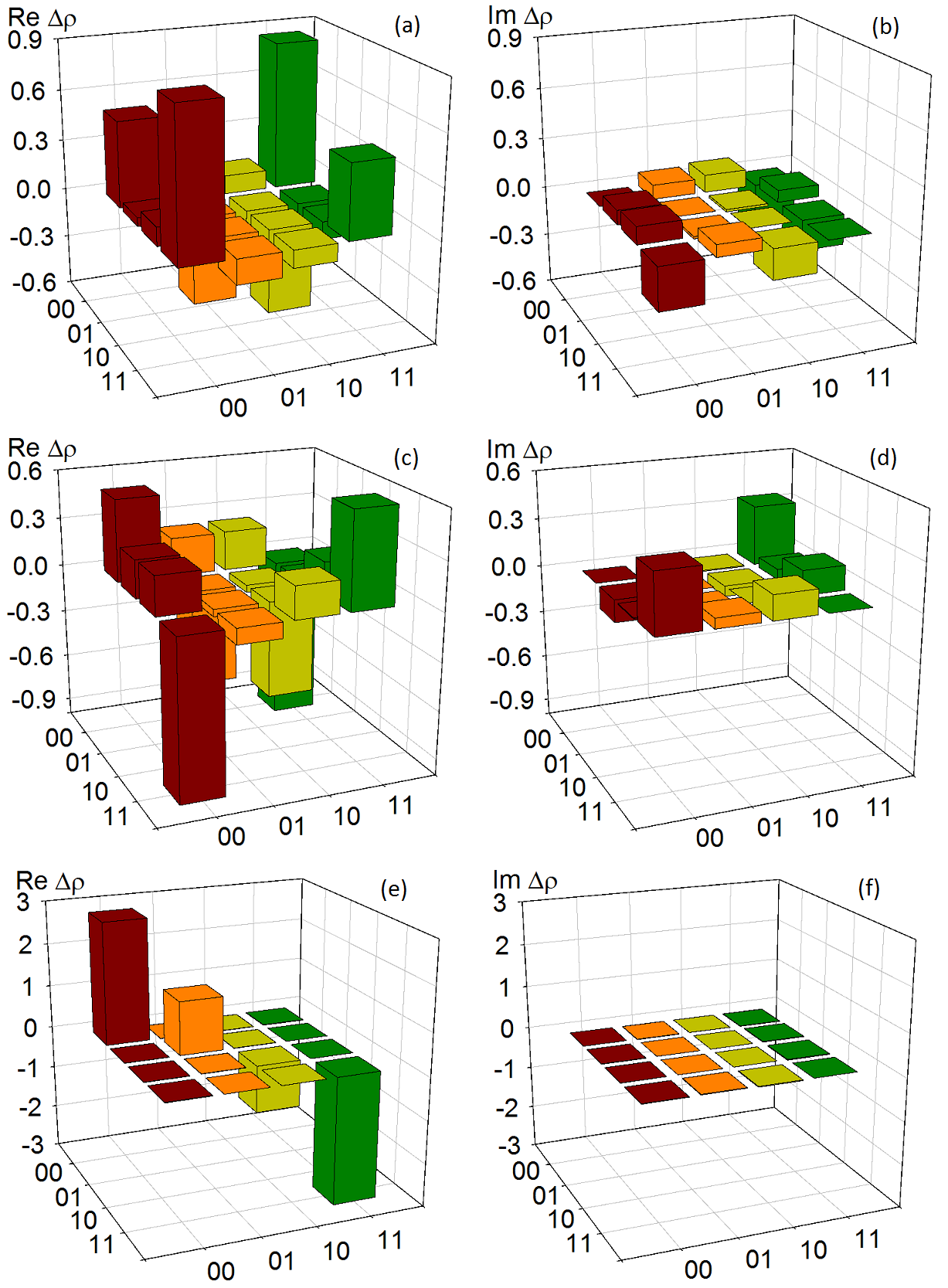}%
\caption{(Color online) Bar representation of the QST experimental data for
the prepared deviation matrix: (a) real and (b) imaginary parts of the
equivalent to the $\left\vert \Psi^{+}\right\rangle $ Bell-basis pseudopure
state; (c) real and (d) imaginary parts of the equivalent to the $\left\vert
\Psi^{-}\right\rangle $ Bell-basis pseudopure state; (e) real and (f)
imaginary parts of the equilibrium state. }%
\end{center}
\end{figure}

In our experiment, the sample of $^{23}$Na nuclei in a lyotropic liquid
crystal, was prepared with $20.9~wt\%$ of SDS ($95\%$ of purity), $3.7~wt\%$
of decanol, and $75.4~wt\%$ of deuterium oxide, by following the procedure
described in Ref. \cite{1976_JPhysChem_80_174}. The $^{23}$Na NMR experiments
were performed in a $9.4$ T -- VARIAN INOVA spectrometer using a $7~mm$
solid-state NMR probe head at $T=26%
{{}^o}%
C$. Figure 2 shows the equilibrium spectrum for our sample. From this spectrum
we obtained the quadrupole frequency $\nu_{Q}=6\omega_{Q}/2\pi=15kHz$ and the
constant parameter $C=(12\pm1)\times10^{9}s^{-2}$. By following the procedure
described in Sec. III C, we estimated the values of the spectral densities as
$J_{0}=(17\pm4)\times10^{-9}s$, $J_{1}=(3.0\pm0.5)\times10^{-9}s$, and
$J_{2}=(3.4\pm0.5)\times10^{-9}s$. Consequently, the time constants that
appear in Eqs. (\ref{Relax1}) and (\ref{Relax2}) are $\tau_{L1}=\left(
13\pm4\right)  $ m$s$, $\tau_{L2}=\left(  14\pm4\right)  $ m$s$, and $\tau
_{T}=\left(  13\pm4\right)  $ m$s$. Finally, the mean values obtained for the
constant parameters $R_{j}$ are $R_{1}=-1.9\pm0.2,$ $R_{2}=-7.9\pm0.1$, and
$R_{3}=-4.5\pm0.2$.%
\begin{figure}
[pt]
\begin{center}
\includegraphics[
trim=0.000000in 1.605395in 0.000000in 1.208184in,
natheight=18.389400in,
natwidth=11.400000in,
height=4.075in,
width=2.9888in
]%
{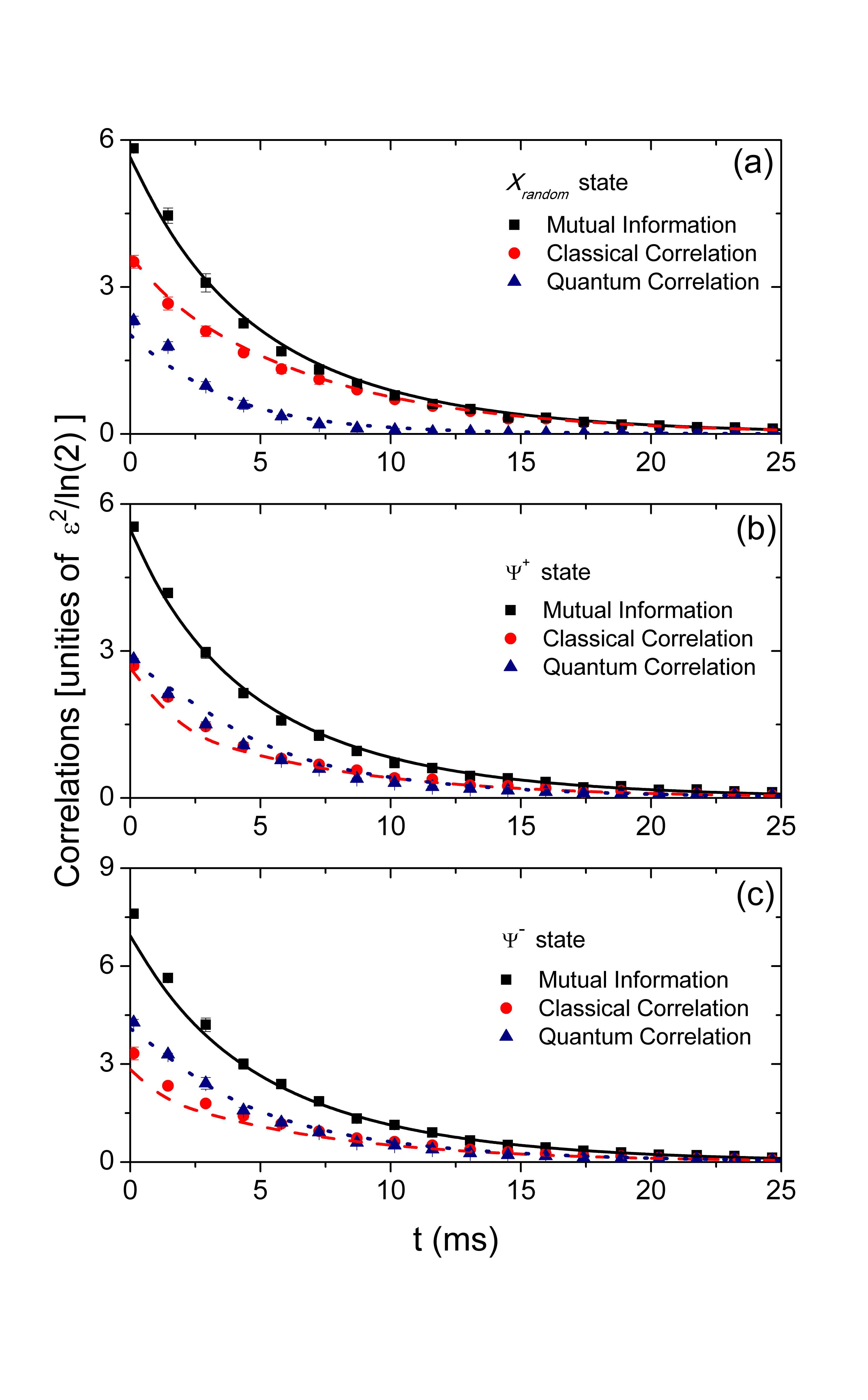}%
\caption{(Color online) Correlations in the effective two-qubit pseudopure
states. The solid lines represent the theoretical model, and the marks are the
experimental points.}%
\end{center}
\end{figure}
\begin{figure}
[pt]
\begin{center}
\includegraphics[
trim=0.000000in 0.384975in 0.000000in 0.772560in,
natheight=13.050000in,
natwidth=11.400000in,
height=3.1185in,
width=2.9914in
]%
{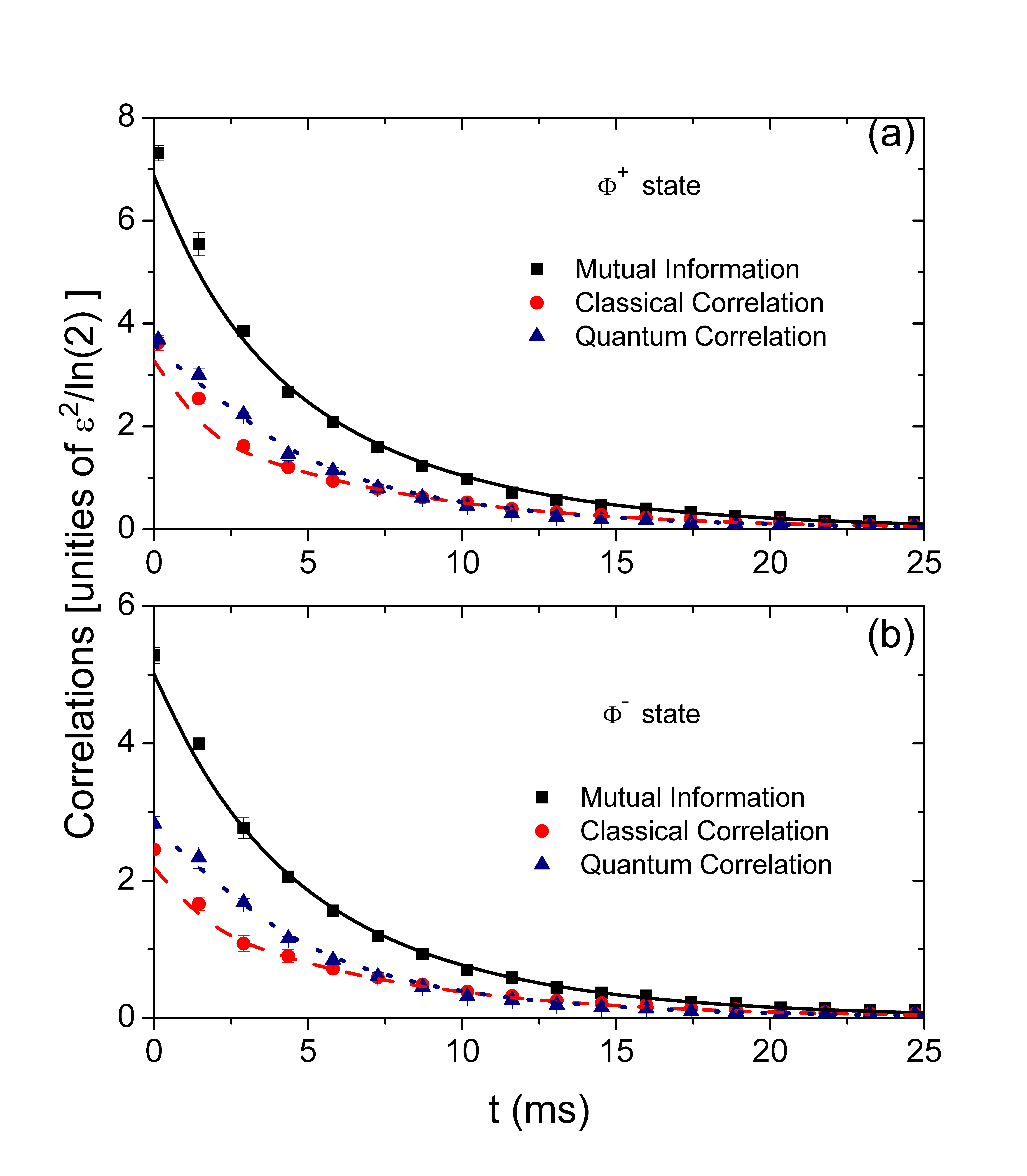}%
\caption{(Color online) Correlations in the effective two-qubit pseudopure
states. The solid lines represent the theoretical model, and the marks are the
experimental points.}%
\end{center}
\end{figure}

By means of numerically optimized rf pulses (SMP technique)
\cite{2002_JChemPhys_116_7599, 2005_JChemPhys_12_2214108,
2007_JChemPhys_126_154506} and temporal averaging, we prepared the
deviation-density matrices, which correspond to five distinct initial sates:
an arbitrary random X-type pseudopure state $\left(  \left\vert
X_{\text{random}}\right\rangle \right)  $ and the four Bell-basis pseudopure
states $\left\{  \left\vert \Psi^{\pm}\right\rangle =\left.  \left(
\left\vert 00\right\rangle \pm\left\vert 11\right\rangle \right)  \right/
\sqrt{2},\left\vert \Phi^{\pm}\right\rangle =\left.  \left(  \left\vert
01\right\rangle \pm\left\vert 10\right\rangle \right)  \right/  \sqrt
{2}\right\}  $. All of these initial pseudopure states have the following
form, in the computational basis, for the corresponding deviation matrices,
\footnote{In our experiment the null elements of the deviation-density matrix
are zero\ within the experimental errors.}%
\begin{equation}
\Delta\rho_{R}=\left[
\begin{array}
[c]{cccc}%
a & 0 & 0 & f\\
0 & b & e & 0\\
0 & e^{\ast} & c & 0\\
f^{\ast} & 0 & 0 & d
\end{array}
\right]  .
\end{equation}
The five prepared initial-state deviation matrices are displayed as a bar
representation in Figs. 4(a)-4(f) and Figs. 5(a)-5(d). The mean fidelity of
the \textit{prepared} states relative to the \textit{ideal }ones is
$\mathcal{F}=\left(  \sqrt{\rho_{\text{ideal}}}\rho_{\text{prepared}}%
\sqrt{\rho_{\text{ideal}}}\right)  ^{1/2}\approx0.98$. We note that,
independent of the initial state considered, the asymptotic state is always
given by the thermal equilibrium state whose deviation-matrix is $\Delta
\rho=2I_{z}$ [displayed in Figs. 5 (e) and 5(f)].

The reconstruction of the deviation-density matrix was performed through QST
described in Ref. \cite{2007_JChemPhys_126_154506}. The relaxation dynamics of
the elements was followed using the procedure described in Sec. III C. We
computed the quantum correlation and its classical counterpart present in the
effective two-qubit\ pseudopure state accordingly Sec. III B [i.e., they were
computed through Eqs. (\ref{SQC}), (\ref{QMI_DM}) and (\ref{CC}) by using the
experimental reconstructed deviation-matrix for different evolution periods
$\tau=n\Delta\tau$ ($n=1,..,40$)]. For each initial state, the free evolution
depicted in the pulse sequence displayed in Fig. 3 was performed 40 times with
an evolution-period increment of $\Delta\tau=1.5$ m$s$, leading to a maximum
evolution time of $60$ m$s$. The errors bars in Figs. 6(a)-6(c) and 7(a) and
7(b) were estimated from the standard deviation of three or more realizations
of the whole experimental procedure (sketched in Fig. 3) for each initial
state (prepared via the SMP technique).

We display the decoherence dynamics of correlations (classical, quantum and
the mutual information) present in the two-qubit pseudopure states
corresponding to $\left\vert X_{\text{random}}\right\rangle ,$ $\left\vert
\Psi^{+}\right\rangle $, $\left\vert \Psi^{-}\right\rangle $, $\left\vert
\Phi^{+}\right\rangle $, and $\left\vert \Phi^{-}\right\rangle $ in Figs.
6(a)-6(c), 7(a), and 7(b), respectively. The theoretical curves obtained from
the model presented in Sec. III C are in very good agreement with the
experimental data. In these figures, we observe that the correlations decay
monotonically (as an exponential law) under the action of the nuclear-spin
environment. We also note that for the Bell-basis pseudopure states
[$\left\vert \Psi^{\pm}\right\rangle ,\left\vert \Phi^{\pm}\right\rangle $,
Figs. 6(b) and 6(c), and 7(a) and 7(b)], the quantum correlation is typically
greater than its classical counterpart $\mathcal{Q}(\rho_{AB})\geq
\mathcal{K}(\rho_{AB})$. The opposite situation occurs for the chosen
$\left\vert X_{\text{random}}\right\rangle $ pseudopure state [Fig. 6(a)]
where $\mathcal{Q}(\rho_{AB})\leq\mathcal{K}(\rho_{AB})$. The former
experimental observation contradicts the early conjecture that the classical
support of the correlation was believed to be equal to or greater than its
quantumness for any quantum state \cite{2005_PRA_71_062307,
2001_JPhysA_34_6899, 2003_PRL_90_050401, 2005_PRA_72_032317}. However, such an
observation is consistent with previous theoretical predictions
\cite{2009_PRA_80_044102, Luo3}. It is also interesting to compare the
dynamics of the quantum correlation with the relaxation of the system as
described in Sec. III C. Note, for example, that the decoherence time of the
quantum correlation for the state $|X_{random}\rangle$ is about $8$ m$s$, the
same order of magnitude for the relaxation characteristic times $\tau_{T}$,
$\tau_{L_{1}}$, $\tau_{L_{2}}\sim13$ m$s$. The others initial pseudopure
states exhibit a similar behavior.

\section{Concluding remarks}

Although entanglement is usually criticized in NMR systems, several quantum
protocols have been successfully implemented and have been tested in such a
scenario \cite{1999_PRL_83_1054, 2001_PRL_87_047901, 2002_QIC_2_166,
livro_NMRivan}. The debate about the quantum nature of NMR implementations for
QIP \cite{2002_QIC_2_166} is renew due to the fact that separable states can
exhibit nonclassical correlation, which may be responsible for the
computational speedup in the NMR context \cite{2009_ArXiv_0906.3656}.
Therefore the investigations about such general correlation in an NMR system
become quite relevant.

In this paper we theoretically and experimentally studied the dynamics of
bipartite quantum and classical correlations in an NMR quadrupolar system
under the action of the decoherence process (mainly caused by the EFG
fluctuations). We reported the first observation of a nonclassical correlation
in an effective two-qubit NMR system. We also provide an approach for
computing the quantum and classical correlations in such a composed system
from the experimentally accessible NMR deviation-density matrix.

We have found that classical and quantum correlations decay monotonically in
time, by following an exponential law, which is in perfect agreement with the
behavior predicted by the theoretical model that was presented. By depending
on the initial state (random or Bell-basis pseudopure states), the relation
between the amount of quantum and classical correlations changes. Our results
show that the nonclassical correlation can be generated and can be manipulated
in spin-pseudopure states of NMR systems at room temperature. In our
experiments, the quantum aspect of correlation decays monotonically within
the\ longitudinal and transverse relaxation characteristic times. Although the
experimental realizations were performed in a specific quadrupolar NMR system,
the methods developed here are quite general and can readily be applied to
other NMR systems or whichever system uses a deviation-density matrix
formalism. It is worth mentioning that, despite the fact that we prepared
specific pseudopure states used in the context of the NMR QIP, the results are
valid for any initial NMR state concerning its specific $\Delta\rho$. In other
words, initial states obtained after simpler excitation schemes without
temporal or spatial averaging may also have nonclassical correlations that
show similar behavior. The results presented here may shed new light on the
quantum resources available in the NMR scenario for QIP implementations.

\begin{acknowledgments}
We acknowledge financial support from the Brazilian funding agencies CAPES,
CNPq and FAPESP. J.M. and L.C.C. are also grateful to UFABC. This work was
performed as part of the Brazilian National Institute for Science and
Technology of Quantum Information (INCT-IQ).
\end{acknowledgments}

\end{document}